\documentclass[aps,prx,reprint,superscriptaddress,twocolumn,showpacs]{revtex4-1}
\setcitestyle{numbers,square}

\usepackage{amsmath}
\usepackage{amssymb}
\usepackage{float}
\usepackage{dcolumn}
\usepackage{bm}
\usepackage{epsf}
\usepackage{graphicx}
\usepackage{amsfonts}
\usepackage{sidecap}

\sidecaptionvpos{figure}{c}

\usepackage[colorlinks=true,citecolor=blue,linkcolor=blue,urlcolor=blue]{hyperref}

\begin{document}

\title{Critical role of magnetic moments in heavy-fermion materials: revisiting mysteries of SmB$_{6}$}

\author{Ruiqi~Zhang}
\affiliation{Department of Physics and Engineering Physics, Tulane University, New Orleans, LA 70118, USA}

\author{Bahadur~Singh}
\affiliation {Department of Condensed Matter Physics and Material Science, Tata Institute of Fundamental Research, Colaba, Mumbai 400005, India}

\author{Christopher~Lane}
\affiliation{Theoretical Division, Los Alamos National Laboratory, Los Alamos, New Mexico 87545, USA}
\affiliation{Center for Integrated Nanotechnologies, Los Alamos National Laboratory, Los Alamos, New Mexico 87545, USA}

\author{Jamin~Kidd}
\affiliation{Department of Physics and Engineering Physics, Tulane University, New Orleans, LA 70118, USA}

\author{Yubo~Zhang}
\affiliation{Department of Physics and Engineering Physics, Tulane University, New Orleans, LA 70118, USA}

\author{Bernardo~Barbiellini}
\affiliation{Department of Physics, School of Engineering Science, Lappeenranta University of Technology, FI-53851 Lappeenranta, Finland}
\affiliation{Department of Physics, Northeastern University, Boston, MA 02115, USA}

\author{Robert S. Markiewicz}
\affiliation{Department of Physics, Northeastern University, Boston, MA 02115, USA}

\author{Arun~Bansil}\email{ar.bansil@neu.edu}
\affiliation{Department of Physics, Northeastern University, Boston, MA 02115, USA}

\author{Jianwei~Sun}\email{jsun@tulane.edu}
\affiliation{Department of Physics and Engineering Physics, Tulane University, New Orleans, LA 70118, USA}

\begin{abstract}
$\newline$
{Heavy-fermion family exhibits fascinating and often puzzling properties due to the presence of open-shell $f$ ions and the complexity of the associated charge, orbital, and spin degrees of freedom. SmB$_6 $ is a prototypical heavy-fermion compound that is electrically insulating but yet it displays quantum oscillations, which are a telltale signature of the metallic state. Adding to the enigma is the possibility that SmB$_6$ is a topological Kondo insulator. Here, by treating the spin degree of freedom on an equal footing with other degrees of freedom using the parameter-free strongly-constrained and appropriately-normed (SCAN) density functional, we explore the ground-state electronic structure of SmB$ _{6}$. A number of competing magnetic phases lying very closely in energy are found, indicating the key role of spin fluctuations in the material. The computed band structure, crystal-field splittings in the $f$-electron complex, the heavy effective electron mass at the Fermi energy, and the large specific heat are all in good agreement with the corresponding experimental results. In particular, our predicted FS explains the experimentally observed bulk quantum oscillations as well as the low electrical conductivity of SmB$ _{6}$. The topological Kondo state of SmB$_6$ is shown to be robust regardless of its magnetic configuration. The excellent performance of SCAN in heavy-fermion systems is explained in terms of its ability to treat self-interaction errors and symmetry breaking within the framework of the density functional theory. Our study provides a new approach for modeling heavy-fermion materials.}
\end{abstract}

\maketitle

\section{Introduction}\label{introduction}

Correlated materials with open $d$- or $f$-shells have long been at the center of condensed matter research theoretically and experimentally due to their fascinating properties~\cite{mott2004metal,Coleman1}, including metal insulator transition~\cite{Imada1998}, non-Fermi liquid behavior~\cite{GR2001}, the Kondo effect~\cite{Heavy2020}, and unconventional superconductivity~\cite{Fradkin2015}. This complexity is driven by the interplay of various degrees of freedom, of which spin is perhaps the most intriguing in that it results in the emergence of competing orders with a rich tapestry of phase diagrams. Local magnetic moments of $d$- and $f$-ions have drawn less attention when long range magnetic ordering is not observed experimentally~\cite{Zhang2020,Varignon2019,Zhang2020b,wang2020PRB}.

In correlated materials involving transition metal elements, we have predicted the presence of ‘intertwined orders’, where the ground state competes with multiple nearly-degenerate phases with strong local magnetic moments~\cite{Zhang2020,Furness2018,Lane2018,Lane2020}. Non-magnetic phases are found not to be energetically competitive, suggesting that local magnetic moments must be present on some disordered or short-range ordered scale. Given the similarities in spin structure and differences in orbital characters of the $d$ and $f$ electrons, with rapidly growing interest in heavy-fermion physics~\cite{Coleman1}, it is important to ask the question: What is the role of local magnetic moment in driving properties of open $f$-shells systems?

We attempt to address this question with the example of SmB$_{6} $ as a prototypical heavy-fermion system. SmB$_{6} $ has been of special interest as a potential strongly correlated topological insulator (TI)~\cite{Li2020NPR}. In fact, its possible surface state was reported over 50 years ago, long before TIs were known~\cite{Menth1969}. Recent classification of SmB$_6 $ as a new class of strongly correlated electrons--a topological Kondo insulator (TKI)~\cite{Dzero2010}, has attracted considerable theoretical~\cite{Lu2013,Kang2015,Chang2015} and experimental~\cite{Li2014,Tan2015a,Hartstein2018,Fuhrman2019,Chen2018,Frantzeskakis2013,Jiang2013,Neupane2013,Dzero2016} interest, although the situation remains ambiguous. Some angle-resolved photoemission spectroscopy (ARPES)~\cite{Frantzeskakis2013,Jiang2013,Neupane2013,Xu2013a}, scanning tunneling spectroscopy~\cite{Roler2014,Jiao2016}, and de Haas-van Alphen (dHvA) oscillation studies~\cite{Li2014,Xiang2017} claim to have found the topological surface states. But other dHvA experiments~\cite{Tan2015a,Hartstein2018,Hartstein2020} indicate that the quantum oscillations (QOs) of SmB$_{6} $ are bulk like.  This conclusion is supported also by recent x-ray Compton scattering experiments~\cite{2021arXiv211107727M} but it is seemingly incompatible with SmB$_{6}$ being an insulator. Moreover, the experimentally measured low-temperature linear specific heat of SmB$ _{6 }$ cannot be attributed to a surface state~\cite{Hartstein2018,Wakeham2016,Phelan2014}. It has been suggested~\cite{Thomas2019} that one of the QOs in SmB$_{6}$ may be an artefact due to aluminum inclusions. These seemingly contradictory results must be understood in order to gain a handle on the strange electronic and topological behavior of SmB$_{6} $.

In this connection, a variety of theoretical models have been proposed. These include a number of exotic~\cite{Baskaran2015,Knolle2017,Erten2017,Chowdhury2018} and disorder-based ~\cite{Shen2018,Harrison2018,Sen2020} models, which are designed to explain the co-existence of QOs and low electric conductivity in SmB$_{6} $. Among the first-principles studies, various flavors of the density-functional theory (DFT)~\cite{Kang2015,Chang2015,Antonov2002,Gmitra2014,Sakhya2020} have not been able to provide valence $f$-band splittings and Fermi surface (FS) topologies of SmB$_{6}$ in accord with experimental results. These fundamental failures are usually ascribed to the difficulty of describing strongly localized $f$ electrons in DFT due to self-interaction errors (SIE)~\cite{Perdew1981} and the competition between the itinerant and localization tendencies~\cite{Strange1999}. Dynamical mean-field theory (DMFT)~\cite{Kim2014} and DFT+Gutzwiller~\cite{Lu2013} have also been applied to SmB$_6 $. Early DMFT calculations did not obtain the correct $f$-band splittings~\cite{Frantzeskakis2013,Neupane2013,Xu2013a}, although recent work~\cite{dmft2} captures $f$-band splittings, insulating response, and the TKI state.  However, since ref~\cite{dmft2} found an insulator, they did not address the problems of QOs and specific heat. It may be noted that DMFT is typically much more expensive computationally compared to DFT.  

Although much of the existing literature on  SmB$ _{6} $  assumes a non-magnetic ground state,  recent experiments demonstrate the presence of localized magnetic moments on Sm sites with short-range magnetic correlations at low temperatures. Notably, the rounded maximum in the magnetic susceptibility observed in XMCD~\cite{Fuhrman2019} and the nonzero average values of the hyperfine interactions found in nuclear forward-scattering experiments~\cite{Barla2005}  confirm the presence of intrinsic short-range magnetic correlations. Moreover, muon-spin-rotation ($ \mu $SR)~\cite{Biswas2014,Gheidi2019} and NMR~\cite{Caldwell2007} find the magnetic fluctuations to be homogeneous throughout the volume of the sample. Magnetic fluctuations obviously require first-principles treatments that go beyond the non-magnetic case~\cite{Zhang2020,Zhang2020b}.

Here, we employ the strongly-constrained and appropriately-normed (SCAN) density functional~\cite{Sun2015} to examine SmB$_{6}$. All degrees of freedom (spin, orbital, charge, and lattice) are treated on an equal footing without invoking any free parameters such as the Hubbard U. We consider several magnetic configurations of SmB$ _{6}$, which lie extremely close in energy. Our calculations capture the correct Sm $f$-band splittings and the large specific heat in good agreement with the corresponding experimental results. We predict bulk FSs composed of hybridized $d$ and $f$ orbitals which are in good agreement with the QO experiments when small effects of Sm vacancies are included. Moreover, we find a convergence of several factors that can contribute to exceptionally low conductivity, including the heavy effective electron mass predicted at the FS. We also reconcile the seemingly incompatible experimental observations related to the existence of topological surface states. Finally, the excellent performance of SCAN is explained by its improved treatment of SIE and the effect of stabilized local magnetic moments.

\section{Methodology}\label{Methods}

All calculations were performed by using the pseudopotential projector-augmented wave method~\cite{Kresse1999} as implemented in the Vienna ab-initio simulation package (VASP)~\cite{Kresse1993,Kresse1996}. A high energy cutoff of 520 eV was used to truncate the plane-wave basis set. The exchange-correlation effects were treated using the strongly-constrained-and appropriately-normed (SCAN) meta-GGA scheme~\cite{Sun2015}. For the energies and unfolded band structure calculations, we adopted a 2 $\times$ 2 $\times$ 2  supercell for all the magnetic structures considered with a 6 $\times$ 6 $\times$ 6 $\Gamma$-centered $k$ mesh to sample the bulk BZs. Spin-orbit coupling effects were included self-consistently. The crystal structures and ionic positions were fully optimized with a force convergence criterion of 0.01 eV/\AA{} for each atom and a total energy tolerance of 10$^{-5} $ eV. For the PM phase, the polymorphous representation of magnetic moments at different Sm sites was implemented with the special-quasi-random structure (SQS) model~\cite{Zunger1990,Wei1990}. We used the stochastic generation algorithm implemented in the alloy theoretic automated toolkit (ATAT)~\cite{VandeWalle2002,VandeWalle2013} code to search for the best SQS for our 56 atom supercell. To explore the FS topology, we adopted a 1 $\times$ 1 $\times$ 2 supercell for the \textit{A}-AFM model with 4225 $k$ points in order to get a high-quality plot. The predicted QO frequencies were calculated using the SKEAF program~\cite{Rourke2012}. We used BandUp to obtain the unfolded band structure~\cite{Medeiros2014,Medeiros2015}. The FS was obtained with the FermiSurfer code~\cite{Kawamura2019}.

\section{Results}\label{results}
\subsection{Crystal, magnetic, and electronic structures}\label{cme}

\begin{figure*}[htbp]
	\begin{center}
		\includegraphics[width=0.8\textwidth]{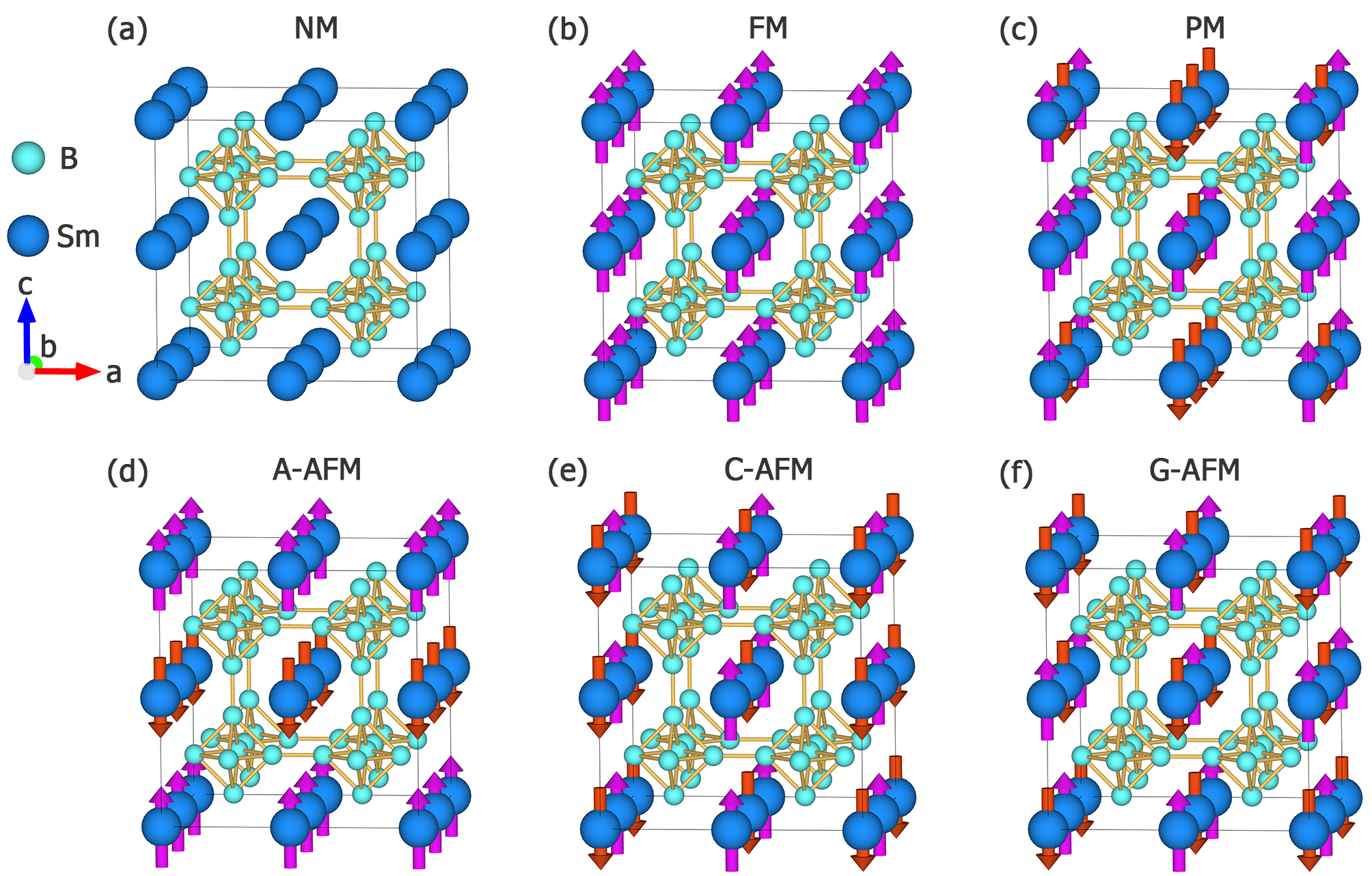}
	\end{center}
	\caption{Structure models for various magnetic configurations. (a) Non-magnetic (NM), (b) ferromagnetic (FM), (c) special-quasirandom paramagnetic structure (SQS-PM), (d) \textit{A}-type antiferromagnetic (\textit{A}-AFM), (e) \textit{C}-type antiferromagnetic (\textit{C}-AFM), and (f) \textit{G}-type antiferromagnetic (\textit{G}-AFM) configurations. The blue and aqua balls represent Sm and B atoms, respectively. The orange and purple arrows denote different magnetic moment directions}
	\label{fig:fig1}
\end{figure*}

SmB$ _{6} $ is known to crystallize in the CsCl-type structure with the Sm atoms located at the corners and B$ _{6 } $ octahedral cluster lying at the body center of the cubic lattice. In order to better understand the nature of the ground state, we consider non-magnetic (NM), paramagnetic (PM), and several magnetic states, including the ferromagnetic (FM) as well as the classical \textit{A}-, \textit{C}-, and \textit{G}-type antiferromagnetic (AFM) configurations. The PM phase is modeled by a special quasi-random structure (SQS)~\cite{Zunger1990,Wei1990} with a 56-atom supercell. We adopt a 2$ \times $2$ \times $2 supercell as shown in Fig.~\ref{fig:fig1}. Table~\ref{Table1} gives the relaxed lattice constants for all magnetic configurations considered, which agree well with the corresponding experimental value of 4.133  \AA{}~\cite{Roler2014}.

\begin{table*}[htbp]
	\setlength{\belowcaptionskip}{0.5cm}
	\renewcommand\arraystretch{1.4}
	\caption{The calculated lattice constants, energies, magnetic moments, and topological characters for various magnetic configurations of SmB$ _{6} $.} 
	\label{Table1}
	\centering
	\tabcolsep 0.6cm
	\begin{tabular}{c|ccc|c|c|c}\hline\hline
		Phase & \multicolumn{3}{c|}{Lattice Constant (\AA{})} & Energy & $M _{Sm} $& Topological \\
		& a \quad &b\quad  &c \quad & (meV/atom) & ($ \mu $B) &  Character \\ \hline
		\textit{A}-AFM	& 4.126 \quad &4.126\quad  &4.129 \quad & 0 & 5.447 &  Non-trivial \\ 
		\textit{SQS}-PM	& 4.128 \quad &4.128\quad  &4.128 \quad & +2.29 & 5.446 &  Non-trivial \\
		FM	& 4.129 \quad &4.129\quad  &4.129 \quad & +3.42 & 5.458 &  Non-trivial \\
		\textit{C}-AFM	& 4.126 \quad &4.126\quad  &4.122 \quad & +5.18 & 5.433 &  Non-trivial \\ 
		\textit{G}-AFM	& 4.129 \quad &4.129\quad  &4.129 \quad & +5.88 & 5.425 &  Non-trivial \\ 
		NM	& 4.119 \quad &4.119\quad  &4.119 \quad & +717.12 & 0 &  Non-trivial \\ 
		\hline\hline
	\end{tabular}
\end{table*}

\begin{figure*}[htbp]
	\begin{center}
		\includegraphics[width=0.9\textwidth]{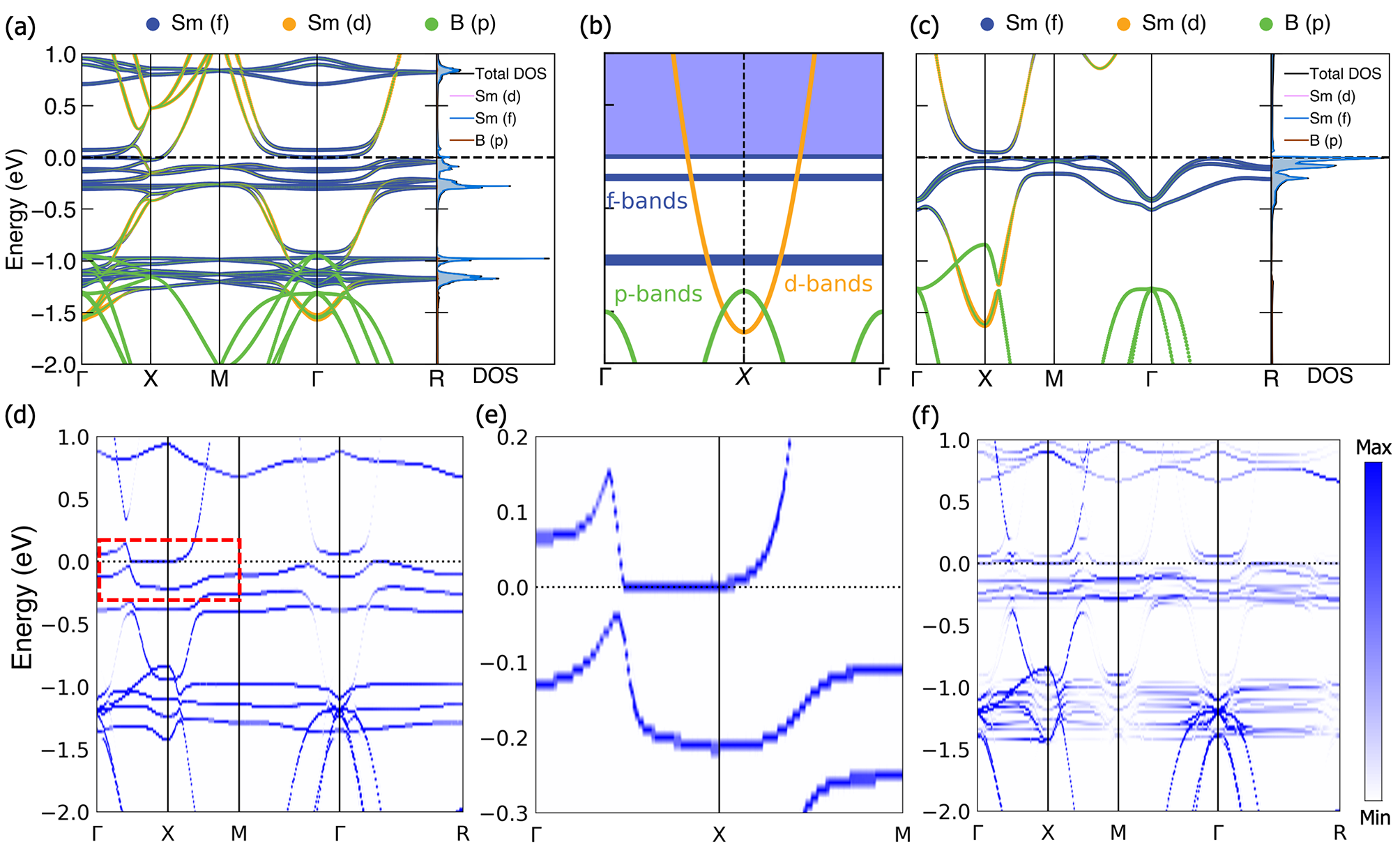}
	\end{center}
	\caption{Bulk band structure of SmB$ _{6 } $  for various magnetic configurations. (a) Orbital-resolved band structure and density-of-states (DOS) of \textit{A}-AFM in the Brillouin zone (BZ) of the $ 2\times2\times2 $ real-space supercell. (b) Schematic illustration of the observed SmB$_6$ band structure in ARPES experiments. Blue, yellow and green lines represent $f$-, $d$- and $p$-states, respectively. f-bands represent those seen in Ref.~\cite{Frantzeskakis2013,Neupane2013,Xu2013a}. (c) Similar to (a), but for the nonmagnetic phase with primitive cell. (d) The unfolded band structure corresponding to (a) in the primitive $ 1\times1\times1 $  BZ. (e) Closeup of the area marked by the red-dashed rectangle in (c) to highlight the flat $f$ band and the $d-f$ hybridized bandgap near  $ E_{F} $. (f) The unfolded band structure for the SQS-PM configuration. The $f$-state degeneracy is now further reduced due to the disordering of spins at the Sm sites. Intensity of the dispersion features in frames (d)-(f) is a measure of their unfolded spectral weight.}
	\label{fig:fig2}
\end{figure*}

As seen from Table~\ref{Table1}, the total energies of different magnetic configurations are very close to one another although the \textit{A}-type AFM (\textit{A}-AFM) phase has the lowest energy. Magnetic moments are $ \sim $5.4 $ \mu_B$ in all cases. Interestingly, the energy and local magnetic moments in the SQS-PM model are similar to those of the \textit{A}-AFM phase. Note that the NM state that has been the focus of much of the earlier work lies at 717.12 meV/atom higher than the \textit{A}-AFM state. Our results emphasize the importance of quantum fluctuations and competing orders in SmB$ _{6} $. These findings are consistent with those on other strongly correlated materials with competing orders~\cite{Zhang2020,Varignon2019,Zhang2020b,Furness2018,Lane2018}.

Next, we present the band structures for the \textit{A}-AFM and SQS-PM phases in Fig.\,\ref{fig:fig2}, the two lowest energy phases, along with that of the energetically-disfavored NM phase. The $f$-electron band complex in \textit{A}-AFM and SQS-PM is seen to be divided into three groups of bands in agreement with ARPES experiments~\cite{Frantzeskakis2013,Neupane2013,Xu2013a}. In contrast, in the NM case the $f$-valence bands stay clustered around the Fermi level  (Fig.\,\ref{fig:fig2} (c)). These results highlight the important role of local magnetic moments in SmB$_6$.

The 2$ \times $2$ \times $2 supercell based band structure of the $A$-AFM state [Fig.\,\ref{fig:fig2}(a)] and its unfolded representation on the original 1$ \times $1$ \times $1 Brillouin zone (BZ) [Figs.\,\ref{fig:fig2}(d) and (e)] show that the three crystal-field-split $f$ states are located around -0.13, -0.30, and -1.0 eV. Similar $f$-band splittings also occur in the SQS-PM [Fig.\,\ref{fig:fig2}(f)] as well as other magnetic states, see Fig. S1 of supplementary materials (SM). As expected, the degeneracy of the $f$ bands in SQS-PM is lower compared to other magnetic states due to the local spin-disorder on the Sm sites. These results are in substantial accord with the ARPES measurements and, to the best of our knowledge, such a level of agreement was only achieved in the latest DFT+DMFT calculations~\cite{dmft2}, but not in earlier studies~\cite{Lu2013,Kang2015,Antonov2002,Gmitra2014,Kim2014}. Note that while ARPES is a surface-sensitive spectroscopy, it can probe both the surface and bulk states with relative sensitivity that depends on the photon energy used in the measurements. SCAN's ability to capture $f$-band splittings is further discussed in Section~\ref{SCAN}

Figure~\ref{fig:fig2} shows that the low-energy states in SmB$ _{6} $ mainly consist of the dispersive Sm 5$d$ and flat 4$f$ bands, irrespective of the magnetic configuration. There is a hybridization gap due to the mixing of the $d$ and $f$ bands away from the high-symmetry points at $ E_{F} $.  The Sm $f$ bands are essentially flat with a high density of states (DOS) of 61.80  states/eV  at $ E_{F} $ [Fig.\,\ref{fig:fig2}(a)]. The associated specific heat coefficient (theoretical) is $ \gamma  $ = 18.21 mJ K$ ^{-2} mol ^{-1} $, which is consistent with the experimental value of 10 $\sim$ 50 mJ K$ ^{-2} mol ^{-1} $~\cite{Hartstein2018,Wakeham2016,Phelan2014}, indicating that the unusual specific heat capacity of SmB$ _{6} $ is a bulk effect. The preceding analysis strongly indicates the critical role of localized magnetic moments in SmB$ _{6} $ for predicting band structure and specific heat. Next, we move to discuss the co-existence QOs and low electric conductivity.

\subsection{Quantum Oscillations}\label{QOs}
\subsubsection{SCAN-based predictions}\label{ourscanQOs}
Although there is a $d-f$ hybridization gap in SmB$ _{6} $, the ground state of \textit{A}-AFM is metallic with extremely flat, heavy-fermion-like bands near the $ E_{F} $  [Fig.\,\ref{fig:fig2} (d)]. We will show below how such a band structure can lead to QOs similar to those observed in experiments~\cite{Hartstein2018,Hartstein2020}. 

\begin{figure*}[htbp]
	\begin{center}
		\includegraphics[width=0.9\textwidth]{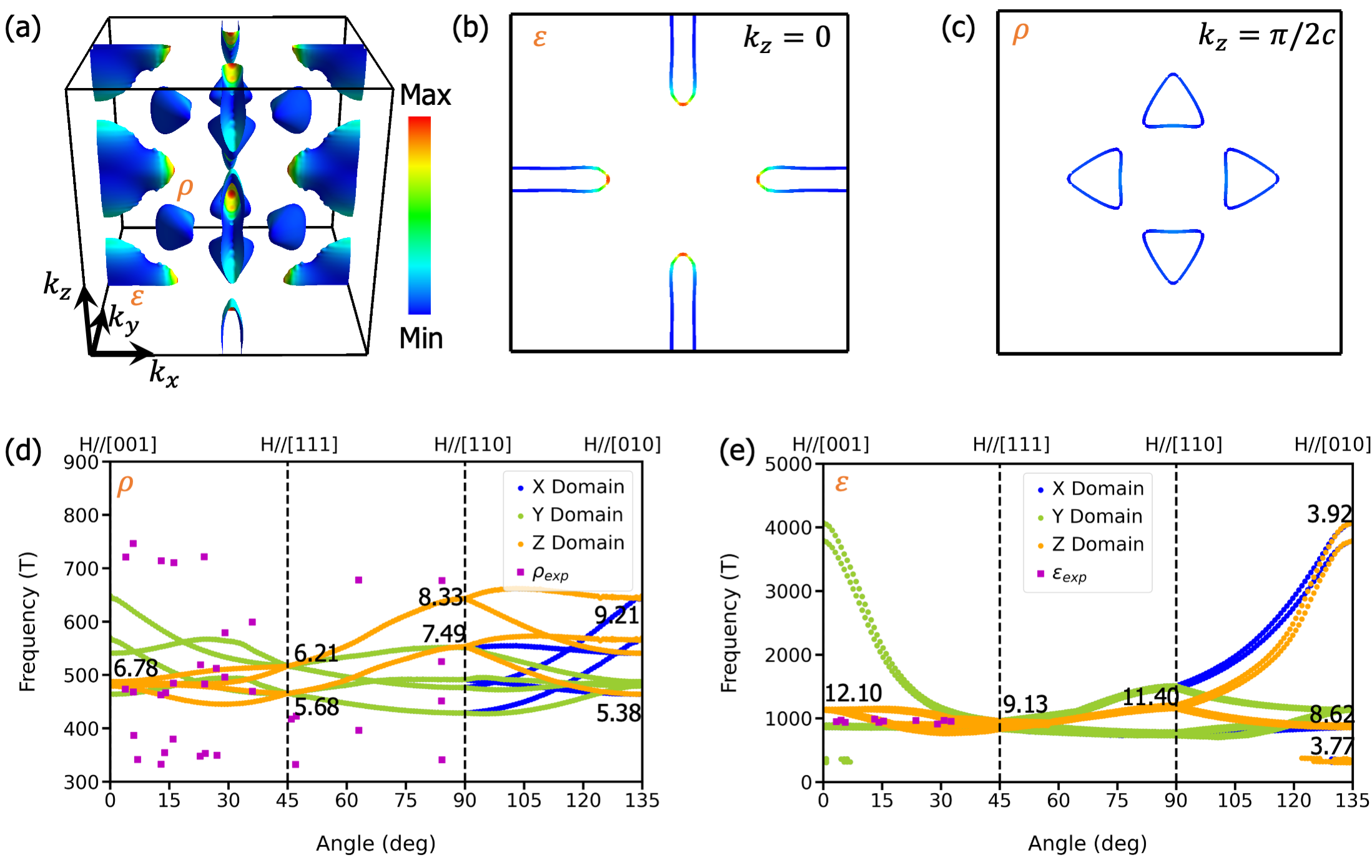}
	\end{center}
	\caption{Theoretical Fermi surface (FS) and related quantum oscillation (QO) frequencies for the unfolded \textit{A}-AFM structure from $ 1\times1\times2  $ superlattice BZ. (a) The unfolded FS for the \textit{A}-AFM in the primitive $ 1\times1\times1  $ BZ. The $ \varepsilon $ and $ \rho $ Fermi pockets are marked. The color code indicates the Fermi velocity. Projection on (b) $k_z=0$ plane of $\varepsilon$ and (c) the $k_z=\pi/2c$ plane of the $\rho$ pockets. (d) Calculated QO frequencies of the $\rho$ pockets with the field rotating from the [001] to [010] direction. Effective masses (m$ ^{*} $/m$ _{e} $) in [001], [111], [110] and [010] directions are marked on the red colored lines. Different color lines indicate different branches associated with X-, Y- or Z- oriented domain wall. (e) Same as (d) but for the $\varepsilon$ Fermi pockets.$\rho_{exp}$ and $\epsilon_{exp}$ electron pockets are from ref~\cite{Hartstein2020}}
	
	\label{fig:fig3}
\end{figure*}

Figure\,\ref{fig:fig3} shows that the unfolded FS of \textit{A}-AFM phase with a $1 \times 1  \times2 $ supercell consists of two distinct sheets with their symmetry-related replicas on the primitive $1 \times 1  \times1 $ BZ. The larger surface centered at the X point near the $k _{z} $ = 0 plane is labeled as $ \epsilon  $ whereas the surfaces lying on the $ k_{z} $= $ \pm \pi/2c $ plane are identified as $\rho$. In Fig. {\color{blue} S2} of SM, we show the evolution of the unfolded FS on the primitive $ 1\times 1  \times 1$ BZ if the Fermi energy is varied. The $\varepsilon$ surface is very sensitive to changes in $ E_{F} $ and its shape changes drastically with just a 1 meV (or 11.6 K) shift of $ E_{F} $ [Fig. {\color{blue} S2}]. Importantly, when we increase $ E_{F} $ by 3 meV, our theoretical results come into reasonable accord with experiments and, for this reason, we will now focus on the results at E= $ E_{F} $+3 meV in Fig.\,\ref{fig:fig3}. Shifts in Fermi energies with respect to the computed band structures are quite commonly invoked in the literature due to uncertainties inherent in first-principles computations and also to account for the effects of doping and defects on the materials involved in the experiments. Moreover, SmB$_6$ lies close to an insulator-metal transition, making it especially susceptible to tuning by applied magnetic field and material parameters. Note also that we find a diverging mass near the original (unshifted) Fermi level due to flat $f$ bands, see Supplementary Section 1 for details.

The calculated QOs for different magnetic field orientations associated with the $\rho$ Fermi pockets are presented in Fig.\,\ref{fig:fig3} (d). Branches resulting from X-, Y- and Z-oriented crystallites are shown to highlight multiplicity of possibilities. The area of this surface varies from 450-650 T which is somewhat narrower than the corresponding experimentally observed value $\rho_{exp}$  of 309-750 T~\cite{Hartstein2018}. We find eight symmetry-related pockets which lie on the $k_z=\pm\pi/{2c}$ planes [Figs.\,\ref{fig:fig3} (a) and (c)]. Notably, experiments not only find these eight pockets but also their symmetry related counterparts on the $k_x=\pm\pi/{2a}$ and $k_y=\pm\pi/{2b}$ planes~\cite{Hartstein2018}. These experimental results can be understood naturally if we include two other $A$-AFM domains with short axes along the $x$ and $y$ directions. There are eight $\rho$ pockets centered at $(\pm\pi/2a,0, \pm\pi/2c)$ and $(0, \pm\pi/2b, \pm\pi/2c)$ for the $A$-AFM domain with short-axis along $z$. Similarly, the Fermi pockets are located at $(\pm\pi/2a, \pm\pi/2b,0)$ and $(\pm\pi/2a,0, \pm\pi/2c)$ for the short axis along the $x$ axis and at $(\pm\pi/2a, \pm\pi/2b,0)$ and $(0, \pm\pi/2b, \pm\pi/2c)$ for the short axis along the $y$ axis. These 24 pockets are degenerate in pairs leaving only 12 pockets in agreement with experiments. Such a multi-domain scenario is commonly observed in experiments where heterogeneous nucleation leads to strong twinning of the allowed domains. In the interest of brevity, we will discuss only our results with the \textit{z}-oriented \textit{A}-AFM domains.

The calculated QO frequencies for the $\varepsilon$ pockets of the FS are shown in Fig.\,\ref{fig:fig3} (e). Interestingly, the $\varepsilon$ pockets, illustrated in Figs.\,\ref{fig:fig3} (a) and (b) display an anomalous `flatfish' shape, where the flat portions come from the Sm $f$ states while the large cross-sections result from the Sm \textit{d} states [Figs.\,\ref{fig:fig2} (a), (d), (e)]. The calculated frequency for the $\varepsilon$ pockets is quite small everywhere, except for fields along the  [001]-[111], and [110]-[010] directions where the frequency approaches the  experimental values ($\alpha_{exp}$)~\cite{Hartstein2018}. 

The experimental QO spectrum can thus be understood as follows. The flat band ($\varepsilon_{exp}$) seen in ref~\cite{Hartstein2018} agrees well with our $\varepsilon$ flatfish band along the [001]-[111]-[110] directions. The $\alpha_{exp}$ pockets seen in the experiments may be the result of a magnetic breakdown, which results in the decoupling of the hybridized $d$ and $f$ bands to yield a purely $d$-like FS. The $\alpha_{exp}^{'}$ band at $\sim$7,000 T may be another breakdown feature that arises from the mixing of the $\varepsilon_{exp}$- and $\alpha_{exp}$-band frequencies. In order to model the effects of the magnetic breakdown, we have repeated our calculations by restoring the unhybridized $d$ bands at the Fermi level by artificially moving the $f$ electrons into the core region, see Section~\ref{CEBS}. The resulting frequencies are comparable to the $\alpha_{exp}$ frequencies of $\sim$8,000 T~\cite{Hartstein2018}. In this way, we are able to account for all the experimentally observed QO features.

We emphasize that the anomalous `flatfish' $\varepsilon$ pockets could explain the insulating character of SmB$_{6} $.  Note first that the large experimental $\alpha_{exp}$ band will make no contribution to transport if it reflects the effects of magnetic breakdown under high fields as we discussed above. The flat $\varepsilon$ band with angle-dependent effective masses with a maximum value of $\sim$21 m$ _{e} $, however, will lead to nearly localized heavy-fermion carriers which can act as strong scatterers [see Fig.~{\color{blue}S4}]. The observed reduction of effective mass with increasing temperature~\cite{Denlinger2013} suggests that the $ f $ electrons rapidly become incoherent and decouple from QOs, which is in keeping with the fact that our zero-temperature effective masses are greater than the experimental values~\cite{Hartstein2018,Thomas2019}. Ref.~\cite{Harrison2018} finds that the electron mass is 0.18 m$_{e}$ at temperatures above 1 K, but rapidly increases to 30 m$_{e}$ at lower temperature, consistent with our calculated values at T=0 K. Since our band structure computations refer to zero temperature, we are not in a position to address temperature effects. We further note that a recent STM study~\cite{Pirie2020} finds effective masses as large as 410$\pm$20 m$_{e}$, consistent with $f$-electron physics. Also, the `flatfish' FS lies close to a Van Hove singularity (VHS) [Fig.\,\ref{fig:fig3} (b)], which could drive strong scattering even at low temperatures and add an anomalously large temperature-dependent correction to the resistivity. Moreover, the VHS could induce a diverging effective mass near the termination of the magnetic order, much like the case of the cuprates near the charge-density-wave critical point~\cite{Ramshaw2015}. Some evidence for the presence of a VHS in SmB$_{6}$ is provided by the appearance of a low-temperature peak in the thermal conductivity measurements~\cite{Hartstein2018}. Correlated materials are known to show anomalous transport due to intertwined orders or neutral FS consisting of itinerant low-energy excitations that can transport heat but not charge~\cite{Hartstein2018,Sodemann2018}. Our results thus make the electrically insulating bulk state in SmB$_6$ more plausible. Transport calculations in this connection will be interesting. Notably, a robust transport gap has been reported experimentally~\cite{Eo2019}.

Since the $\rho$ and $\epsilon$ FS shift in opposite directions with doping [see Fig.~{\color{blue}S2}], there is a unique doping at which their average areas match experiment.  However, we find that the shape-anisotropy of the calculated $\rho$ FS is smaller than that in experiment.  This may be related to a little-known paradox of $f$-electron physics.  As temperature rises and the $f$-electrons become incoherent, one expects them to cease contributing to the FS, leading to a transition from a small ($f-d$ hybridized) to a large ($d$-only) FS. However, what is often seen instead is a transition from a large-mass to a small-mass FS with no change in area, as discussed by Harrison in SmB$_6$~\cite{Harrison2018}. This suggests that there is an intermediate phase, where the FS is still small, but the mass and shape of the FS are controlled by $d$-electrons only, which could explain the difference in shape of the observed $\rho$ pockets at high $T$.

\begin{figure*}[htbp]
	\begin{center}
		\includegraphics[width=0.85\textwidth]{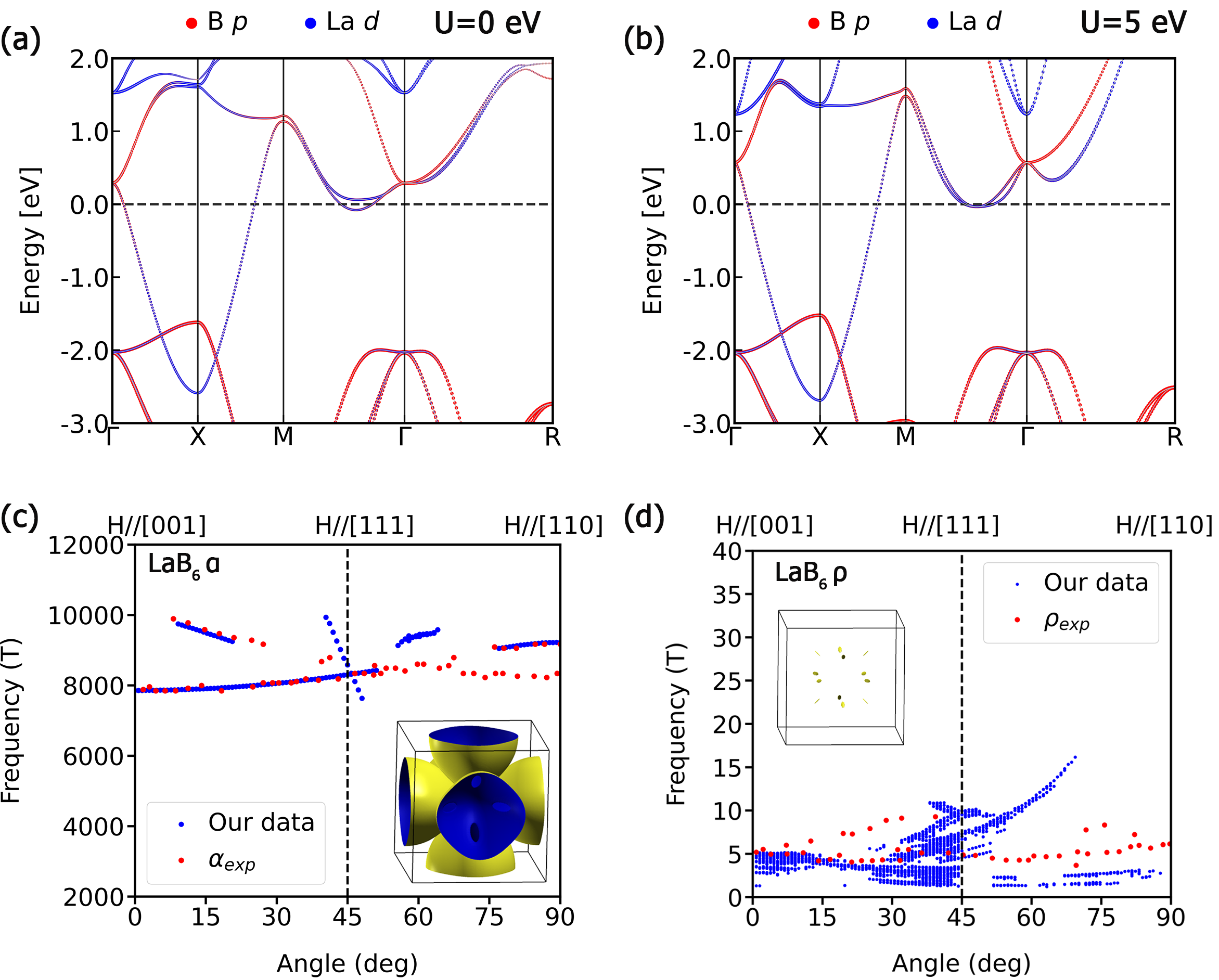}
	\end{center}
	\caption{(a)-(b) Calculated band structure of LaB$_{6}$ using SCAN and SCAN+U (U=5 eV), respectively. The two bands near $E=0$ along the branch with $k$ lying between the $M$ and $\Gamma$ points play a crucial role in forming the $\rho$-pocket in LaB$_6$.  When the Fermi level crosses the lower band, adjacent $\alpha$-ellipsoids overlap and open a hole pocket in the overlap region.  The $\rho$-pocket develops from the tips of the adjacent $\alpha$-ellipsoids in this hole when the Fermi level crosses the upper band.  Computed QO frequencies for $\alpha$ (c) and $\rho$ (d) Fermi pockets of LaB$_{6}$ based on SCAN+U. The applied magnetic field is rotated from [001] to [010] direction. $\alpha_{exp}$ and $\rho_{exp}$ electron pockets are from Ref.~\cite{Hartstein2020}.}
	\label{fig4}
\end{figure*}

\subsubsection{Previous Models}\label{PM}

\begin{figure*}[htbp]
	\begin{center}
		\includegraphics[width=0.85\textwidth]{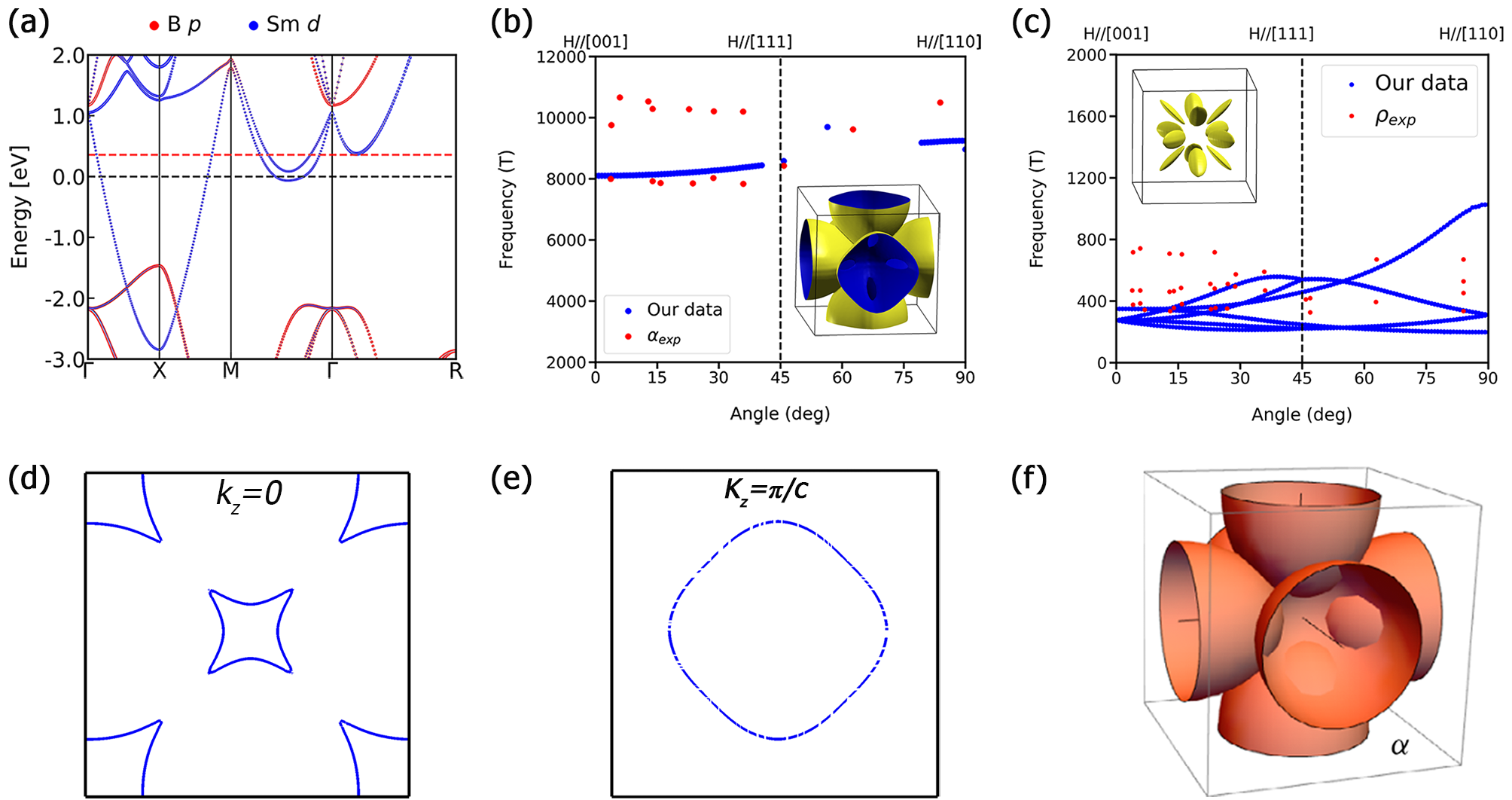}
	\end{center}
	\caption{(a) Simulated band structure of Sm$^{nof}$B$_6$ in which $f$ electrons are artificially removed. Black and red dashed lines represent $E _{F} = 0$ and $E _{F} = 360$ meV, respectively. (b) Calculated QO frequencies for the $\alpha$ Fermi pockets at \textit{E$ _{F}$}. (c) Calculated QO frequencies for the $\rho$ Fermi pockets at $E _{F} = 360$ meV. The applied magnetic field in (b) and (c) is rotated from the [001] to the [110] directions. Experimental $\alpha_{exp}$ and $\rho_{exp}$ data is taken from ref.~\cite{Hartstein2018}. The calculated $\alpha$ and $\rho$ pockets are also shown. (d) and (e) show the $\alpha$ Fermi pocket projections on $k_{z}=0$ and  $k_{z}=\pi/c$ plane, respectively. (f) $\alpha_{exp}$ electron pockets from ref~\cite{Hartstein2020}}
	\label{fig5}
\end{figure*}

There has been considerable debate about the origin of QOs in SmB$_{6}$ from both experiment and theory.  In experiments, ref~\cite{Li2014} reported a two-dimensional (2D)-like FS with a low frequency (less than 1000T), which was suggested to be due to topological surface states, whereas ref~\cite{Thomas2019} argues that these QOs might be due to aluminum inclusions. More recent studies~\cite{Hartstein2018,Hartstein2020}, however, found the low frequency but also a high frequency over 10,000 T. This high frequency agrees with one found in LaB$_{6}$ and with our calculations when the $f$ electrons of Sm were not considered (see Figs.~\ref{fig4} and~\ref{fig5}), consistent with a 3D FS expected for bulk metallic SmB$_6$. The comparison of QOs in SmB$_6$ and in aluminum for a wide frequency range rules out the contribution from aluminum ~\cite{Hartstein2020}. The fact that the low-frequency oscillation develops an $f$-electron contribution below 1 K~\cite{Harrison2018} also rules out the aluminum interpretation. Refs.~\cite{Hartstein2018,Hartstein2020} also report possible signatures of 2D QOs in a narrow range of angles near $\theta = 0$. As we show later, since SmB$_6$ is topologically nontrivial, one cannot rule out the possibility of contributions from surface states.

To explain the 3D FS observed in the electrically insulating SmB$_{6}$, two contrasting models have been proposed: (1) The QOs arise conventionally~\cite{Shen2018,Harrison2018}, with the gaps closed via a combination of thermal and disorder broadening, albeit with disorder strong enough to explain the low conductivity but weak enough to avoid washing out the QOs. (2) The QOs are exotic~\cite{Baskaran2015,Knolle2017,Erten2017,Chowdhury2018}, arising from charge-neutral FSs that do not contribute to electrical transport.  The fact that bulk SmB$_6$ was recently found to be metallic~\cite{Harrison2018} would seem to point in favor of conventional theories.

The Cambridge group found a series of dHvA frequencies in SmB$_{6}$~\cite{Tan2015a,Hartstein2018,Hartstein2020}, which are similar to those for LaB$_{6}$, i.e. where only $d$-electrons are involved, and include a large cluster of ellipsoids that form the $\alpha_{exp}$-bands. They thus argued that a conventional interpretation of the QOs would predict that SmB$_{6}$ is a good metal, contrary to experiment, leading them to propose exotic models of spinless fermions. However, this explanation suffers from the problem that the $d$-bands appear to account for all the FS. If so, there is no room left for strong $f-d$ coupling effects at the Fermi energy, which are a characteristic feature of a Kondo insulator. This model therefore cannot account for the observed hybridization gap~\cite{Frisk2007}, which takes the material from large to small FS.  $f$ electrons are also a necessary ingredient of the spinless fermion model~\cite{Chowdhury2018}  invoked by the Cambridge group for SmB$_6$.  Specifically, in the spinless fermion model~\cite{Chowdhury2018}, the presence of  $f$ electrons is required as they are the source of the spinon-holon pairs. Other exotic theories~\cite{Baskaran2015,Knolle2017,Erten2017} have been proposed to explain QOs of SmB$_6$. In contrast, Harrison~\cite{Harrison2018} argues that $f$-electrons are present in SmB$_{6}$ but they only show up below $\sim$1 K as an anomalously strong increase of the dHvA amplitude: this is a conventional model, which involves Sm vacancies instead of spinons and holons and makes a natural connection with our $T=0$ results as discussed further below.

\subsubsection{Comparison with early band structures}\label{CEBS}

We now compare our calculations with the early non-magnetic band structures that underlie the exotic model interpretation of the Cambridge group ~\cite{Tan2015a, Hartstein2018,Hartstein2020}, who also present QOs of LaB$_{6}$ for reference. We have also calculated the QOs of LaB$_{6}$ in Fig.~\ref{fig4} using SCAN. Fig.~\ref{fig4} shows that we obtain $\alpha$-bands, absence of $f$ bands near the Fermi level, and the absence of the $\rho$-pocket, all in good accord with earlier findings~\cite{Harima1988}. Following ref~\cite{Harima1988}, in order to restore the $\rho$-pocket, we added U= 5 eV on the $f$ electrons, which effectively shifts the Fermi level with respect to the conduction bands along M to $\Gamma$. This shift produces a small $\rho$-pocket and the $\alpha$-bands [see Fig.~\ref{fig4}], which are in good agreement with both theoretical~\cite{Harima1988} and experimental results~\cite{Hartstein2020}.

Turning to SmB$_6$, we have simulated the band structure for an artificial case where the $f$-electrons are excluded (referred to as Sm$^{nof}$B$_6$ hereon). Sm$^{nof}$B$_6$ is representative of magnetic breakdown of the $f-d$ hybridization, which leads to purely $d$-like FSs. $f$-electrons in Sm$^{nof}$B$_6$ are shifted to core states and yield no $f$-character in the bands and produce FSs [Fig.~\ref{fig5}] very similar to those of LaB$_{6}$. Interestingly, Sm$^{nof}$B$_6$ yields QO frequencies $\sim$8,000 T in agreement with the observed high-frequency pockets of SmB$_6$. Comparing frames (b) and (f) of Fig.~\ref{fig5}, we see that the calculated FSs is very similar to the experimental one~\cite{Hartstein2018} except for one difference. The ellipsoidal FSs assumed in ref~\cite{Hartstein2018} pass through one another without interaction [see Fig.~\ref{fig5} (f)].  In contrast, in our case, the intersecting regions annihilate, leaving holes behind [see Fig.~\ref{fig5} (b)]. Therefore, to recover the experimental branch at $\sim$10,000 T would require another magnetic breakdown to restore the overlapping FSs along the long axis of the ellipsoids. Note that ref.~\cite{Tan2015a} invoked a large shift of the Fermi level above the $f$-electron bands to restore the $\alpha$ pocket in their DFT calculations. Their results agree with ours, in that the $\alpha$-pockets do not overlap enough to produce a $\rho$-pocket, and they were forced to seek an alternative origin for their $\rho$-pockets.

The preceding considerations resolve one of the puzzles of SmB$_6$. If SmB$_6$ contains the same large $\alpha_{exp}$ Fermi pocket as LaB$_{6} $, then how can it have a significantly smaller conductivity than LaB$_6$?  Our analysis above with Sm$^{nof}$B$_6$ reveals the illusory nature of this pocket in that it is due entirely to high-field magnetic breakdown. This pocket is absent in the low-field region due to $d-f$ hybridization, which gaps the FS and results in higher resistivity in SmB$_{6}$ in accord with experiments. However, the $\rho$-pocket is absent in Sm$^{nof}$B$_6$ and a very large shift is needed to restore this pocket~\cite{Tan2015a, Hartstein2018,Hartstein2020}, a shift which is about 100 times larger than that in LaB$_6$ [see Figs.~\ref{fig4} (d) and ~\ref{fig5} (c)]. In fact, even a shift of 360 meV is not adequate for this purpose.  In contrast, our SCAN-based SmB$_6$ results (with $f$ electrons) only needs a 3 meV shift of the Fermi level as shown in Fig.~\ref{fig:fig3}, negligible in comparison with the aforementioned 360 meV shift. A recent study~\cite{2021arXiv211103758L} reports QOs in heat capacity, confirming their bulk origin, and also evidence for strong spin fluctuations below $T^*\sim$15K, which rule out any role of Al inclusions, and determine effective masses considerably larger than those from the Cambridge group ($m^*/m\sim$ 4.5-6.6 vs 0.1-1).  All these results are consistent with our model.  Note that one possibility for erroneous effective masses is magnetic breakdown, which will cause anomalous amplitude variations. We note that there is an alternative interpretation of the $\alpha$ FSs that cannot be ruled out.  If the $f-d$ hybridization is spatially inhomogeneous, the sample could be a `patch map' of hybridized and unhybridized regions, where coherent $f$-electrons exist only in the former regions.  A low-field experiment, where magnetic breakdown is absent, could presumably resolve between the two scenarios.

With fluctuating magnetic configurations, strong $f$-electron participation, and strongly hybridized $f-d$ bands, our SCAN results [Sec.~\ref{ourscanQOs}] provide a viable explanation for the observed QOs, and a more reasonable starting point for a theory of exotic spinon-holon effects than the Cambridge non-magnetic model for the following reasons. (\romannumeral1) Spinons and holons were originally discovered as emergent excitations of a 1D AFM system, and their existence normally requires an AFM background. In the 2D cuprates, spinons and holons have not been found despite much effort~\cite{anderson1997theory}. The best place to look for these quasiparticles would be in frustrated magnets or spin liquids. It is plausible that spinons and holons in 2D will be associated with fluctuating AFM/stripe phases, which occur naturally in our model. Notably, we have identified an emergent spin liquid in the cuprates~\cite{Markiewicz2017}. (\romannumeral2) Alternatively, it has been suggested that spinons and holons are sensitive to disorder, and that the insulating state in SmB$ _{6}$ is more likely associated with a Slater insulator~\cite{Sen2020}, which is the case in our model.

\subsection{Topological Structure}\label{TS}

\begin{figure}[htbp]
	\begin{center}
		\includegraphics[width=0.48\textwidth]{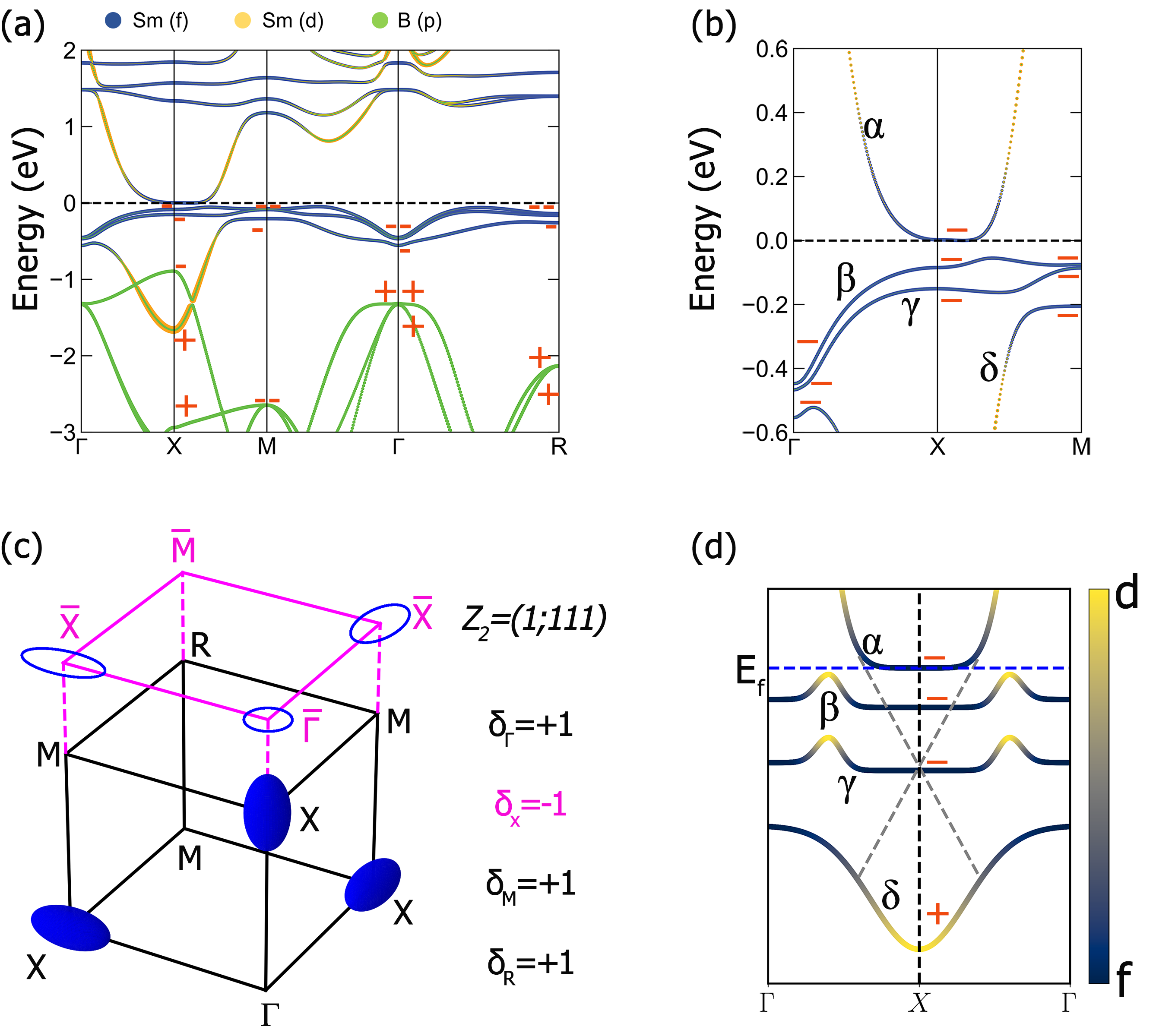}
	\end{center}
	\caption{Topological state of SmB$_6$. (a) Band structure for the NM phase in the primitive ($ 1\times1\times1 $) unit cell. Parity eigenvalues of the occupied bands at the TRIM points are marked. (b) A closeup view of bands along the $\Gamma$-X-M symmetry line in (a). (c) Parity ($ \delta_i $) of the occupied bands at the eight TRIM points and the associated $ Z_{2 } $ invariants. (d) A schematic band structure which is representative of the various magnetic configurations and the associated parity eigenvalues. The expected topological surface states are depicted by dashed lines.}
	
	\label{fig:fig6}
\end{figure}

We turn now to discuss the topological structure of SmB$ _{6} $. The low-energy states in SmB$ _{6} $ are dominated by Sm 5\textit{d} and 4$f$ orbitals, where the 5\textit{d}-derived bands show an exceptionally large degree of itinerancy with a total bandwidth of $\sim$3.0 eV in all of the magnetic structures we have considered. Although the Sm 5\textit{d} states of an Sm$ ^{2+} $ ion are expected to be empty due to their large bandwidth in SmB$ _{6}$, these bands span across $ E_{F} $ and open a robust (immune to magnetic order) inverted hybridization gap at $E_{F}$. Interestingly, we have found that the band structures of the various magnetic states of SmB$ _{6}$ are  adiabatically connected  to  the NM band  structure~\cite{Lin2010}. Therefore, we only need to analyze the topology of the NM band structure. Accordingly, we replot the NM band structure in Figs.\,\ref{fig:fig6} (a) and (b), and compute the topological invariants $ Z_{2} $ = ($ \upsilon_{0} $; $ \upsilon_{1} $$ \upsilon_{2} $$ \upsilon_{3} $), which are well defined for systems respecting time-reversal and inversion symmetries~\cite{Fu2007}. The calculated parity eigenvalues for occupied bands are marked in Figs.\,\ref{fig:fig6} (a) and (b), while the parity of the occupied manifold at each time-reversal invariant momentum (TRIM) point ($ \delta_i $ with $ i $=$ \Gamma $, $ X $, $ M $, and $ R $) is shown in Fig.\,\ref{fig:fig6} (c). The parity is seen to be inverted at the three X points, leading to a TKI state with $ Z_{2} = (1;111) $ in agreement with the previously reported results~\cite{Lu2013,Kang2015}. Note that the inverted parity of the occupied manifold at the X points is preserved in all the magnetic configurations [Fig.\,\ref{fig:fig6} (d)]. The TKI state is thus very robust in SmB$_{6}$, although the associated nontrivial surface states will be sensitive to the details of various magnetic configurations~\cite{Hao2019,Mong2010}.

\subsection{Why does SCAN work in $f$-electron systems?}\label{SCAN}
SCAN~\cite{Sun2015} is a meta-generalized-gradient approximation (metaGGA) to the exchange-correlation (XC) energy functional. Note that DFT is a formally exact theory for the ground state energy and electron density, although the XC component must be approximated for practical calculations. SCAN has been shown to be more accurate than the popular Perdew-Burke-Ernzerhof (PBE) GGA~\cite{PBE} for a wide variety of materials \cite{Sun2015,Sun2016}. SCAN has several advantages over PBE: (1) In addition to the electron density and its gradient used in PBE, SCAN also involves the kinetic energy density. This enables SCAN to satisfy all 17 known exact constraints on the XC energy that a metaGGA can satisfy~\cite{Sun2015}. The 11 exact constraints satisfied by PBE are a subset of the 17 exact constraints. (2) The kinetic energy density is orbital dependent and thus a nonlocal functional of electron density. Therefore, SCAN is typically implemented in the generalized Kohn-Sham (gKS) scheme, where the effective potential is orbital dependent, rather than multiplicative as is the case in the KS scheme. It has been proved that the gKS frontier orbitals have physical meanings for solids and the associated band gap is a physically-justified prediction of the experimentally-measured fundamental band gap \cite{Perdew2017a, Zhang2020b}. But, other orbitals remain auxiliary in gKS. (3) SCAN has less SIE than PBE, which is important for describing open-shell $d$-electron \cite{Zhang2020b} and $f$-electron compounds as discussed below.

\begin{figure*}
	\begin{center}
		\includegraphics[width=0.85\textwidth]{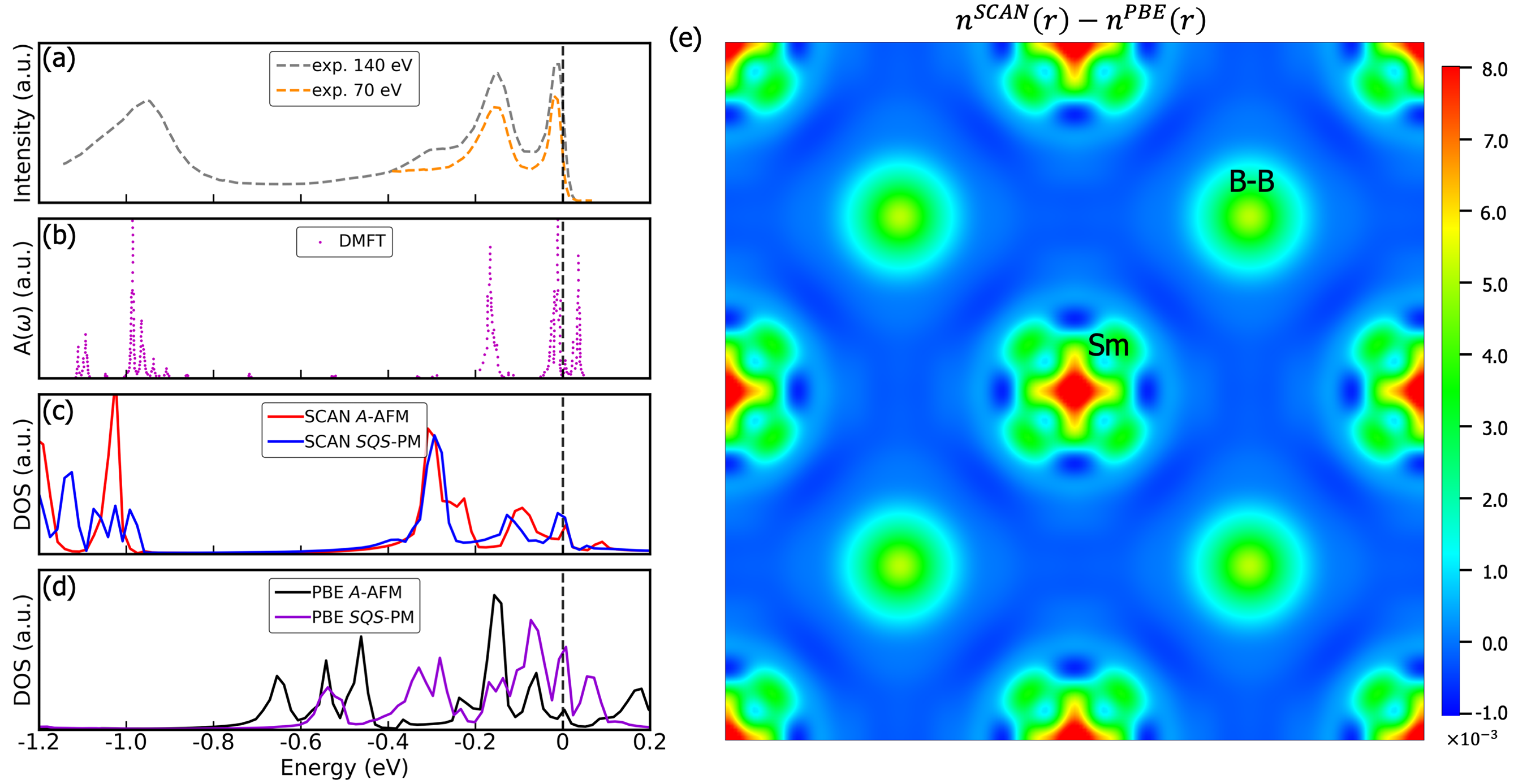}
	\end{center}
	\caption{Various theoretical DOS results for bulk SmB$_{6}$ and the experimental photoemission data.
		(a)  Angle-integrated photoemission spectrum of SmB$_{6}$ adapted from ref.~\cite{dmft2} (b) $k$-integrated spectrum of SmB$_{6}$ from DFT+DMFT adapted from ref.~\cite{dmft2} (c) SCAN and (d) PBE DOS for the $A$-AFM and SQS-PM phases of SmB$_{6}$. (d) DOS of non-magnetic phase of SmB$_{6}$  using SCAN. (e) Difference of electron density calculated by SCAN and PBE for the $A$-AFM phase with a $2\times2\times2$ supercell. The plane is perpendicular to the (001) direction and cuts through Sm atoms and the centers of B-B bonds between two Boron octahedrons. }
	
	\label{fig:fig7}
\end{figure*}

Figs.~\ref{fig:fig7} (a-d) compare the PBE and SCAN based DOSs of SmB$_6$ with the recent DFT+DMFT~\cite{dmft2} and the experimental photoemission results~\cite{Denlinger2014}. SCAN clearly captures split $f$-band peaks at the Fermi energy, -0.13 eV, and -1.0 eV in both the $A$-AFM and SQS-PM configurations, in close agreement with the experimentally observed peaks at the Fermi energy, -0.15 eV, and -0.9 eV, respectively. This corroborates SCAN's good performance discussed in connection with Fig.~\ref{fig:fig2} above. SCAN's predictions here are comparable to those of DFT+DMFT. In contrast, PBE clusters all the $f$-states toward the Fermi energy, with no states below -0.8 eV. Notably, SCAN predicts an additional strong peak around -0.3 eV, where the experimental angle-integrated spectrum ($\hbar{v}$ = 140 eV) shows a broad shoulder that is barely seen in the DFT+DMFT result. We note that SCAN predicts considerable DOS at the Fermi energy, in agreement with the experimental data, while DFT+DMFT yields no states at the Fermi energy. 
 
The improvement of SCAN over PBE discussed above, see also Refs. \cite{Sun2015, Sun2016}, reflects the power of satisfying exact constraints in constructing XC functionals. We emphasize that the improvement in SmB$_{6}$ directly results from the SIE reduction in SCAN compared to the PBE, as is the case also in transition-metal monoxides~\cite{Zhang2020b}. SIE results from the imperfect cancellation of the spurious classical Coulomb self-interaction by the approximate XC functionals. Because the repulsive self-interaction Coulomb energy exceeds the attractive self-XC energy, the net SIE is generally positive. This causes orbitals to be under bound with too high orbital energies and wavefunctions and electron densities to be excessively delocalized. SCAN's lower orbital energies than PBE's in Figs.~\ref{fig:fig7} (c) and (d) indicate clearly that the SIE is smaller in SCAN compared to PBE. Figs.~\ref{fig:fig7}(e) plots the charge density difference between SCAN and PBE for the $A$-AFM phase of SmB$_{6}$ with a $2\times2\times2$ supercell and shows that SCAN localizes more electrons around the Sm atom than PBE by depleting the interstitial electrons. Furthermore, the SCAN electron density around the Sm atom is more anisotropic than PBE's, reflecting the electron localization with respect to the spherical symmetry. The reduced anisotropy of PBE is consistent with its $f$-orbital occupation numbers being more fractional (Table S1), which is another strong indication of the presence of a larger SIE and delocalization errors \cite{Perdew1982, Cohen2008a,Zhang2019c}.

Involvement of the kinetic energy density in SCAN likely does not contribute substantially to the improved description of SmB$_{6}$ in SCAN over PBE. This is because the SCAN-L functional yields a DOS for occupied orbitals which is similar to SCAN (Fig. S5), even though SCAN-L de-orbitalizes SCAN by replacing its kinetic-energy-density dependence with an electron-Laplacian dependence~\cite{SCANL}. SCAN-L is generally more computationally efficient than SCAN, but it violates some exact constraints satisfied by SCAN.

The SCAN-based DOS of the NM state has been shown in Fig.~\ref{fig:fig2} (c), where the $f$-states are clustered towards the Fermi energy more than in PBE A-AFM and SQS-PM with no states below -0.6 eV and are in poor agreement with experimental results. These results again point to the key role of magnetic moments in open-shell $f$-electron compounds, and highlight the importance of spin-symmetry-breaking DFT solutions in correlated systems~\cite{Perdew2021c}. Similar behavior is seen in bulk band structures of other rare-earth hexaborides RB$_6$ (R = Ce, Pr, Nd, Sm, Eu, Gd, Dy and Ho) with FM and NM configurations using SCAN and PBE, see Supplementary Figs. {\color{blue}S6} and {\color{blue}S7}. We note that there have been studies, which show that SCAN overestimates the magnetic moments of bulk transition metals (e.g., bcc Fe)~\cite{DavidPRL} where valence $s$ and $d$ electrons are itinerant. This, however, would not be a concern here since magnetism arises from localized electrons.

We emphasize that most DFT orbitals are auxiliary quantities, with no established formal connection with the physically relevant quasi-particle excitations in materials. The DFT orbitals, however, seemingly capture some underlying physics when an accurate XC functional is used with small SIE and the spin symmetry is allowed to break. In any event, we would expect improved DFT orbitals to provide a better starting point for beyond DFT calculations, including DFT+DMFT and the many-body perturbation methods.

\section{Summary and Conclusions}\label{discussion}

We have systematically examined the electronic and magnetic structures of SmB$_{6}$ using the SCAN density functional without invoking any free parameters such as the Hubbard $U$. Many magnetic phases are found to lie very closely in energy, indicating the propensity of SmB$_{6}$ to harbor competing magnetic orders and spin fluctuations. Our first-principles computations, which involve full self-consistency in charge, spin and lattice degrees of freedom, yield magnetic ground states with crystal-field split $f$ bands in substantial agreement with photoemission results~\cite{Frantzeskakis2013,Neupane2013,Xu2013a, Denlinger2014}. In contrast, the energetically disfavored non-magnetic phase totally misses the crystal-field driven $f$-band splittings. The specific heat is also predicted to be in reasonable accord with the corresponding measurements. The efficacy of SCAN in handling SmB$_6$ is shown to reflect the ability of SCAN to properly localize $f$ electrons by reducing the SIE. Our analysis also gives insight into the surprisingly large bandwidth of Sm 5$d$ states and their hybridization with 4$f$ states and supports the presence of a mixed-valence ground state with Kondo physics in SmB$_6$. Band inversion is shown to occur at the $X$ points irrespective of the magnetic configuration, indicating the robustness of the TKI state in SmB$_6$.

We show that the predicted FS of the ground-state magnetic phase yields bulk QO frequencies in substantial accord with the corresponding experimental results on SmB$_6$. Although the anomalous QOs found in SmB$_6$ have spurred great interest in searching for QOs in other insulators~\cite{Xiang2018a,Sato2019,Wang2021,Xiang2021}, our analysis indicates that there is no single major factor that is responsible for the insulating behavior of SmB$_6$, and that it arises through an interplay that involves break-up of the $d$-electron FSs by $f-d$ hybridization and magnetic order, large $f$-electron effective mass and its divergence caused by the proximity of a flat-band Van Hove singularity, and the fluctuations induced by competing magnetic phases, all of which conspire to reduce the conductivity. The magnetic breakdown that we predict should be amenable to experimental verification through the field dependence of the QO amplitudes. Should this breakdown destroy $f-d$ hybridization and restore the $\alpha$ FSs, it may turn SmB$_6$ from an insulator to a conventional metal. Such an effect has recently been observed in YbB$_{12}$~\cite{Xiang2021}. It would be interesting to ascertain if the breakdown field is related to the 4~meV activation energy reported in SmB$_6$~\cite{Eo2019}.

In summary, our study shows that stabilizing local magnetic moments is the key for resolving puzzling and seemingly contradictory electronic properties of the prototypical heavy-fermion compound SmB$_6$. This result is in line with similar earlier findings in transition metal compounds
~\cite{Zhang2020,Varignon2019,Zhang2020b,wang2020PRB,Furness2018,Lane2018,Lane2020}. Our work thus not only sheds new light on the highly debated mysteries of SmB$_6$, but it also opens a new pathway for simulating compounds with open $f$-shells more generally.

\section*{ACKNOWLEDGMENTS}

The work at Tulane University was supported by the start-up funding from Tulane University, the Cypress Computational Cluster at Tulane, the Extreme Science and Engineering Discovery Environment (XSEDE), the DOE Energy Frontier Research Centers (development and applications of density functional theory): Center for the Computational Design of Functional Layered Materials (DE-SC0012575), the DOE, Office of Science, Basic Energy Sciences Grant DE-SC0019350, and the National Energy Research Scientific Computing Center. The work at Northeastern University was supported by the US Department of Energy (DOE), Office of Science, Basic Energy Sciences Grant No. DE-SC0022216 (modeling complex magnetic states in materials) and benefited from Northeastern University’s Advanced Scientific Computation Center and the Discovery Cluster and the National Energy Research Scientific Computing Center through DOE Grant No. DE-AC02-05CH11231. The work at Los Alamos National Laboratory was supported by the U.S. DOE NNSA under Contract No. 89233218CNA000001 and by the Center for Integrated Nanotechnologies, a DOE BES user facility, in partnership with the LANL Institutional Computing Program for computational resources. Additional support was provided by DOE Office of Basic Energy Sciences Program E3B5.\\

\bibliography{Ref}

\end{document}